\documentclass[pdflatex,sn-mathphys-num]{sn-jnl}% Math and Physical Sciences Numbered Reference Style
\usepackage{graphicx}%
\usepackage{multirow}%
\usepackage{amsmath,amssymb,amsfonts}%
\usepackage{amsthm}%
\usepackage{mathrsfs}%
\usepackage[title]{appendix}%
\usepackage{xcolor}%
\usepackage{textcomp}%
\usepackage{manyfoot}%
\usepackage{booktabs}%
\usepackage{algorithm}%
\usepackage{algorithmicx}%
\usepackage{algpseudocode}%
\usepackage{listings}%
\usepackage{subfig}
\theoremstyle{thmstyleone}%
\theoremstyle{thmstyletwo}%

\theoremstyle{thmstylethree}%

\begin{document}

\title[Quantum Walks–Based Adaptive Distribution Generation with Efficient CUDA-Q Acceleration]{Quantum Walks–Based Adaptive Distribution Generation with Efficient CUDA-Q Acceleration}


\author*[1,2]{\fnm{Yen Jui} \sur{Chang}}\email{aceest@cycu.edu.tw}

\author[3]{\fnm{Wei-Ting} \sur{Wang}}\email{aroe01325@gmail.com}

\author[4]{\fnm{Chen-Yu} 
\sur{Liu}}\email{d10245003@g.ntu.edu.tw}

\author[5]{\fnm{Yun-Yuan} \sur{Wang}}\email{yunyuanw@nvidia.com}

\author[1,2,3]{\fnm{Ching-Ray} \sur{Chang}}\email{crchang@phys.ntu.edu.tw}

\affil*[1]{\orgdiv{Master Program in Intelligent Computing and Big Data}, \orgname{Chung Yuan Christian University}, \orgaddress{\street{No. 200, Zhongbei Rd., Zhongli Dist}, \city{Taoyuan City}, \postcode{320314},  \country{Taiwan}}}
\affil[2]{\orgdiv{Quantum Information Center}, \orgname{Chung Yuan Christian University}, \orgaddress{\street{No. 200, Zhongbei Rd., Zhongli Dist}, \city{Taoyuan City}, \postcode{320314},  \country{Taiwan}}}
\affil[3]{\orgdiv{Department of Physics}, \orgname{National Taiwan University}, \orgaddress{\street{No. 1, Sec. 4, Roosevelt Rd.},\city{Taipei}, \postcode{106319}, \country{Taiwan}}}

\affil[4]{\orgdiv{Graduate Institute of Applied Physics}, \orgname{National Taiwan University}, \orgaddress{\street{No. 1, Sec. 4, Roosevelt Rd.},\city{Taipei}, \postcode{106319}, \country{Taiwan}}}

\affil[5]{\orgdiv{NVIDIA AI Technology Center}, \orgname{NVIDIA Corp}, \orgaddress{\city{Taipei}, \postcode{610101}, \country{Taiwan}}}


\abstract{We present a novel Adaptive Distribution Generator (ADG) that leverages a quantum walks–based approach to generate high precision and efficiency of target probability distributions. Our method integrates variational quantum circuits with discrete-time quantum walks(DTQWs)—specifically, split-step quantum walks(SSQWs) and their entangled extensions—to dynamically tune coin parameters and drive the evolution of quantum states towards desired distributions. This enables accurate one-dimensional probability modeling for applications such as financial simulation and structured two-dimensional pattern generation exemplified by digit representations (0–9). Implemented within the CUDA-Q framework, our approach exploits GPU acceleration to significantly reduce computational overhead and improve scalability relative to conventional methods. Extensive benchmarks demonstrate that our Quantum Walk–Based Adaptive Distribution Generator (QWs-based ADG) achieves high simulation fidelity and bridges the gap between theoretical quantum algorithms and practical high-performance computation.}

\keywords{Quantum Computing, Split-Step Quantum Walks, Entangled Quantum Walks, Adaptive Distribution Generation, CUDA-Q, Variational Quantum Circuits, Generative Modeling, Quantum State Preparation}



\maketitle

\section{Introduction}\label{sec1}

Quantum computing promises a paradigm shift in solving complex computational problems by leveraging inherently quantum phenomena such as superposition, entanglement, and interference \cite{Feynman82,Aharonov93, Nielsen2000, Childs09}. 

These capabilities have spurred significant interest in applying quantum algorithms across a broad spectrum of domains — including optimization \cite{Farhi14}, simulation \cite{Lloyd96,Farhi98}, cryptography \cite{Bennett84,Shor94}, and machine learning \cite{Benedetti19,VenegasAndraca08,ChenDRL2020,ChenSVM2024,ChenMaze2024,Liu2024NN}. Efficient quantum state preparation lies at the heart of these endeavors. As illustrated in Equation~\ref{Eq:eq1},
\begin{equation}
|\psi\rangle = \sum_{i} \sqrt{p_{i}} \;|i\rangle,
\label{Eq:eq1}
\end{equation}
a quantum state is expressed as a linear combination of basis states weighted by the square roots of probabilities. In quantum state preparation, the objective is to engineer a state \(|\psi\rangle\) such that, upon measurement, the outcome probabilities \(p_{i} = |\langle i|\psi\rangle|^2\) match a desired target distribution. In this regard, efficient state preparation inherently generates distribution by carefully adjusting the amplitudes \(\sqrt{p_i}\) via controlled quantum operations. For example, Grover and Rudolph \cite{Grover:loading2002} proposed a scheme in which an ancilla register performs a controlled rotation with angles \(\theta_i\) to produce a superposition state:
\begin{equation} \label{eq:controlratation}
\sqrt{p_i}\,|i\rangle \;\rightarrow\; \sqrt{p_i}\,|i\rangle \otimes \Bigl(\cos\theta_i\,|0\rangle + \sin\theta_i\,|1\rangle\Bigr).
\end{equation}
Using ancilla qubits reduces the circuit depth, and complexity is mitigated at a sub-exponential scale \cite{Zhang2022QSP}. Recent proposals in quantum state loading and generative adversarial networks further demonstrate the feasibility of these methods \cite{Rocchetto2018QSP,Zoufal:qGAN2019,PhysRevLett.129.230504,Iaconis2024}.

Parallel to these developments, variational quantum circuits (VQCs) have emerged as powerful tools for learning and reproducing complex, high-dimensional target distributions \cite{Benedetti19,Cerezo21}. These quantum distribution generators bridge the gap between abstract theoretical models and practical applications by enabling quantum systems to emulate intricate data patterns.

At the same time, quantum walks(QWs)—the quantum analogs of classical random walks—have emerged as robust tools for simulating dynamical processes and generating detailed probability distributions. In discrete-time quantum walks (DTQWs) and their split-step variants (SSQWs), introducing a coin Hilbert space affords enhanced control over interference effects and state evolution. In SSQWs, the coin operation is partitioned into distinct sub-steps, enabling fine-tuning of the underlying quantum dynamics necessary for precise distribution generation \cite{Matsuzawa20,VenegasAndraca08}. Our previous work \cite{Changetal} demonstrated that SSQWs can effectively capture complex financial phenomena and prepare intricate quantum states; however, as these models scale to replicate increasingly sophisticated target distributions, they encounter significant performance bottlenecks.

Recent advances in high-performance computing—such as GPU acceleration via frameworks like CUDA-Q \cite{Kim23}—offer promising avenues to overcome these limitations. We propose a Quantum Walk–Based Adaptive Distribution Generator (QWs-based ADG) that integrates variational quantum circuits with quantum walk–based components to address them. The QWs-based ADG dynamically tunes the coin parameters implemented as three-parameter unitary gates by minimizing the discrepancy between the simulated and target distributions. This adaptive process enhances convergence and fidelity, ensuring that the generated distributions accurately match the desired profiles.

Moreover, by harnessing entanglement between coin registers, our framework naturally extends to two-dimensional generative tasks. We demonstrate this capability by generating digit patterns (0--9) on an \(8 \times 8\) grid, thereby highlighting the model’s potential to capture complex spatial correlations.

The remainder of this paper is organized as follows: Section~\ref{sec:methodology} details the methodology behind the QWs-based ADG and its integration with quantum walk–based circuits on the CUDA-Q platform; Section~\ref{sec:results} presents simulation results and performance benchmarks for one-dimensional distribution modeling as well as two-dimensional pattern generation; and Section~\ref{sec:conclusion} concludes with a discussion of our findings and directions for future research.

\section{Methodology}
\label{sec:methodology}

This section presents our integrated framework that combines variational quantum circuits (VQCs) with a quantum walk–based approach to construct a QWs-based ADG. By harnessing the complementary strengths of VQCs and QWs, our method dynamically learns and reproduces target probability distributions with high fidelity, thereby enabling efficient quantum state preparation and advanced generative modeling. A classical optimizer iteratively updates the quantum circuit parameters, ensuring the system accurately approximates complex distributions. In the following subsections, we first review the fundamentals of QWs and variational quantum circuits. We then describe how these models are integrated within the CUDA-Q platform for enhanced performance. Finally, we illustrate the extension of our approach to two-dimensional pattern generation via entangled QWs.

\subsection{Quantum Walks-Based Approach}
QWs are the quantum analogues of classical random walks and serve as a foundational tool for simulating quantum dynamics and designing quantum algorithms \cite{Aharonov93,Childs09,VenegasAndraca08}. They are broadly categorized into continuous-time quantum walks and DTQWs.

In continuous-time quantum walks, the state evolution is governed by the time-dependent Schrödinger equation:
\begin{equation}
    |\psi(t)\rangle = e^{-iHt}|\psi(0)\rangle,
\end{equation}
where \( H \) is the Hamiltonian of the system and \(|\psi(0)\rangle\) is the initial state.

In contrast, DTQWs incorporate an additional coin degree of freedom which enables non-trivial interference effects. The Hilbert space for a DTQW is given by:
\begin{equation} \label{eq:HilbertSpace}
    \mathcal{H} = \mathcal{H}_{c} \otimes \mathcal{H}_{p},
\end{equation}
where \( \mathcal{H}_{c} \) is the coin Hilbert space with basis \( \{\,|\uparrow\rangle,\,|\downarrow\rangle\} \) and \( \mathcal{H}_{p} \) is the position Hilbert space with basis \( \{|x\rangle : x \in \mathbb{Z}\} \). A typical DTQW evolution at each time step consists of a coin operation followed by a conditional shift:
\begin{equation}
    |\psi(t+1)\rangle = \hat{S}\,\bigl(I \otimes \hat{C}\bigr)\,|\psi(t)\rangle,
\end{equation}
with the coin operator \( \hat{C}(\theta, \phi, \lambda) \) defined as:
\begin{equation} \label{eq:Coinoperator}
    \hat{C} (\theta, \phi, \lambda) =
       \begin{pmatrix} 
       \cos\left(\frac{\theta}{2}\right) & -e^{i\lambda}\sin\left(\frac{\theta}{2}\right) \\[1mm]
       e^{i\phi}\sin\left(\frac{\theta}{2}\right) & e^{i(\phi+\lambda)}\cos\left(\frac{\theta}{2}\right)
       \end{pmatrix},
\end{equation}
and the shift operator given by:
\begin{equation} \label{eq:shiftoperator}
  \hat{S} = |\downarrow\rangle \langle\downarrow| \otimes \sum_{x} |x-1\rangle \langle x| + |\uparrow\rangle \langle\uparrow| \otimes \sum_{x} |x+1\rangle \langle x|.
\end{equation}
An initial state is typically prepared as a product state:
\begin{equation} \label{eq:initialstate}
   |\Psi_0\rangle = (\alpha\,|\uparrow\rangle + \beta\,|\downarrow\rangle) \otimes |x=0\rangle,
\end{equation}
with \(|\alpha|^2 + |\beta|^2 = 1\). The evolution over \( t \) steps is given by:
\begin{equation} \label{eq:QWevolve}
  |\Psi(t)\rangle = \Bigl[\hat{S}\,\bigl(I\otimes\hat{C}\bigr)\Bigr]^t |\Psi_0\rangle = \hat{W}^t|\Psi_0\rangle.
\end{equation}

Split-step quantum walks (SSQWs) further refine this approach by partitioning the coin operation into two sub-steps, leading to an evolution operator of the form:
\begin{equation}
    \hat{W} = S^- \, C_{\vec{\theta_2}} \, S^+ \, C_{\vec{\theta_1}},
\end{equation}
where \( C_{\vec{\theta_1}} \) and \( C_{\vec{\theta_2}} \) are independently tunable coin operators, and \( S^+ \) and \( S^- \) are the corresponding shift operators \cite{Matsuzawa20}. This modularity offers enhanced control over interference effects and state evolution, making SSQWs particularly suited for precise distribution generation.

In the SSQWs framework, the overall evolution is decomposed into two sequential shift operations. The positive shift operator \( \hat{S}^+ \) and the negative shift operator \( \hat{S}^- \) are defined as follows:

\begin{equation} \label{eq:shiftplus}
\hat{S}^+ = |\uparrow\rangle \langle\uparrow| \otimes \sum_{x} |x+1\rangle\langle x| + |\downarrow\rangle \langle\downarrow| \otimes \sum_{x} |x\rangle\langle x|,
\end{equation}
which shifts the walker one step to the right when the coin state is \(|\uparrow\rangle\) and leaves the position unchanged for \(|\downarrow\rangle\).

Similarly, the negative shift operator is defined as:
\begin{equation} \label{eq:shiftminus}
\hat{S}^- = |\uparrow\rangle \langle\uparrow| \otimes \sum_{x} |x\rangle\langle x| + |\downarrow\rangle \langle\downarrow| \otimes \sum_{x} |x-1\rangle\langle x|,
\end{equation}
which shifts the walker one step to the left when the coin state is \(|\downarrow\rangle\) and leaves the position unchanged for \(|\uparrow\rangle\).

Together, these operators allow the SSQWs to adjust the walker’s position based on the state of the coin, enabling enhanced control over the interference and distribution generation in quantum simulations.

\begin{figure}[ht]
    \centering
    \subfloat[\textbf{Conceptual Illustration of SSQWs Evolution.} 
   ]{
        \includegraphics[width=0.4\linewidth]{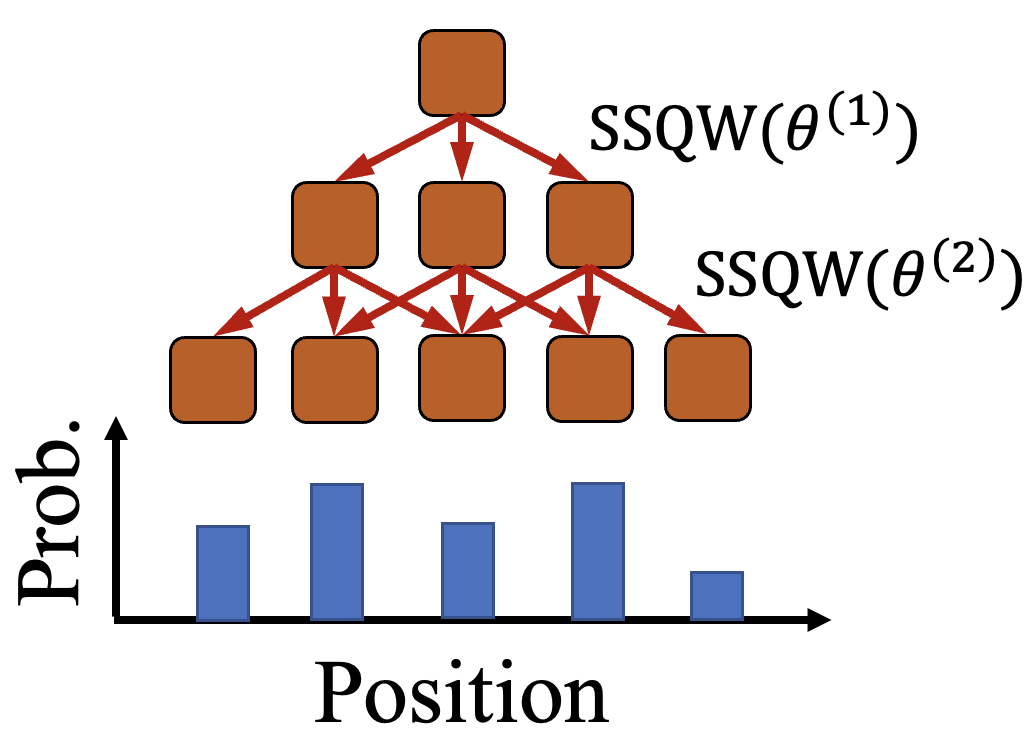}
        \label{fig:ssqw_concept}
    }
    \hfill
    \subfloat[\textbf{Parameterized SSQWs Framework.}
    ]{
        \includegraphics[width=0.45\linewidth]{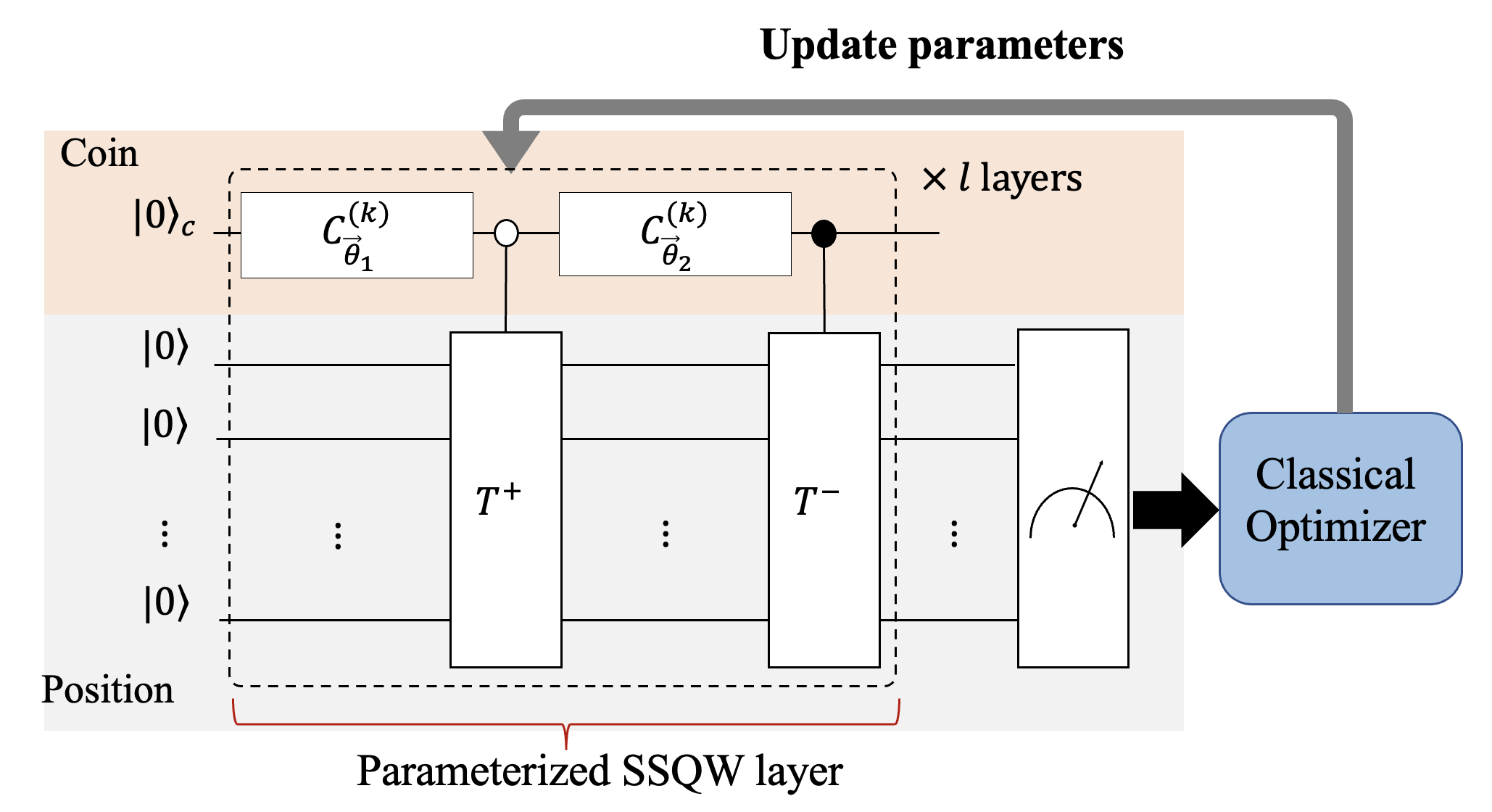}
        \label{fig:ssqw_param}
    }
    \caption{
    Two representations of the SSQWs concept and framework. 
    In (a), the SSQW is depicted as a branching process over multiple layers, illustrating how coin parameters guide the walker’s position probabilities. 
    In (b), a parameterized SSQW layer is shown, where two coin operations ($C_{\vec{\theta_1}}$ and $C_{\vec{\theta_2}}$) and shift operators ($T^+$ and $T^-$) evolve the quantum state, and a classical optimizer iteratively refines the coin parameters to achieve a desired probability distribution.
    }
    \label{fig:ssqw_illustrations}
\end{figure}
Figure~\ref{fig:ssqw_illustrations} encapsulates the core principles of our QWs-based ADG framework. Figure~\ref{fig:ssqw_illustrations}(a) presents a conceptual illustration of SSQWs as a branching process, reminiscent of a trinomial tree, where distinct coin operations sequentially guide the walker’s evolution, thus shaping the probability distribution across successive layers. Figure~\ref{fig:ssqw_illustrations}(b) details a parameterized SSQWs circuit, in which two coin operations, \(C_{\vec{\theta_1}}\) and \(C_{\vec{\theta_2}}\), are interleaved with conditional shift operators (\(T^+\) and \(T^-\)). In this representation, a classical optimizer iteratively refines the coin parameters, implemented as three-parameter unitary gates, to minimize the discrepancy between the simulated and target probability distributions.

Together, these visualizations illustrate how our framework harnesses the adaptive capabilities of variational quantum circuits and the structured evolution of QWs. When accelerated via the CUDA-Q platform, this integration enables efficient and high-fidelity distribution generation, bridging the gap between theoretical quantum models and practical quantum computing applications.

\subsection{Two Entangled Quantum Walkers with Coin Operators in an Entangled Coin Space}
\label{subsec:entangled_QWs}

\begin{figure}[ht]
    \centering
    \includegraphics[width=0.95\linewidth]{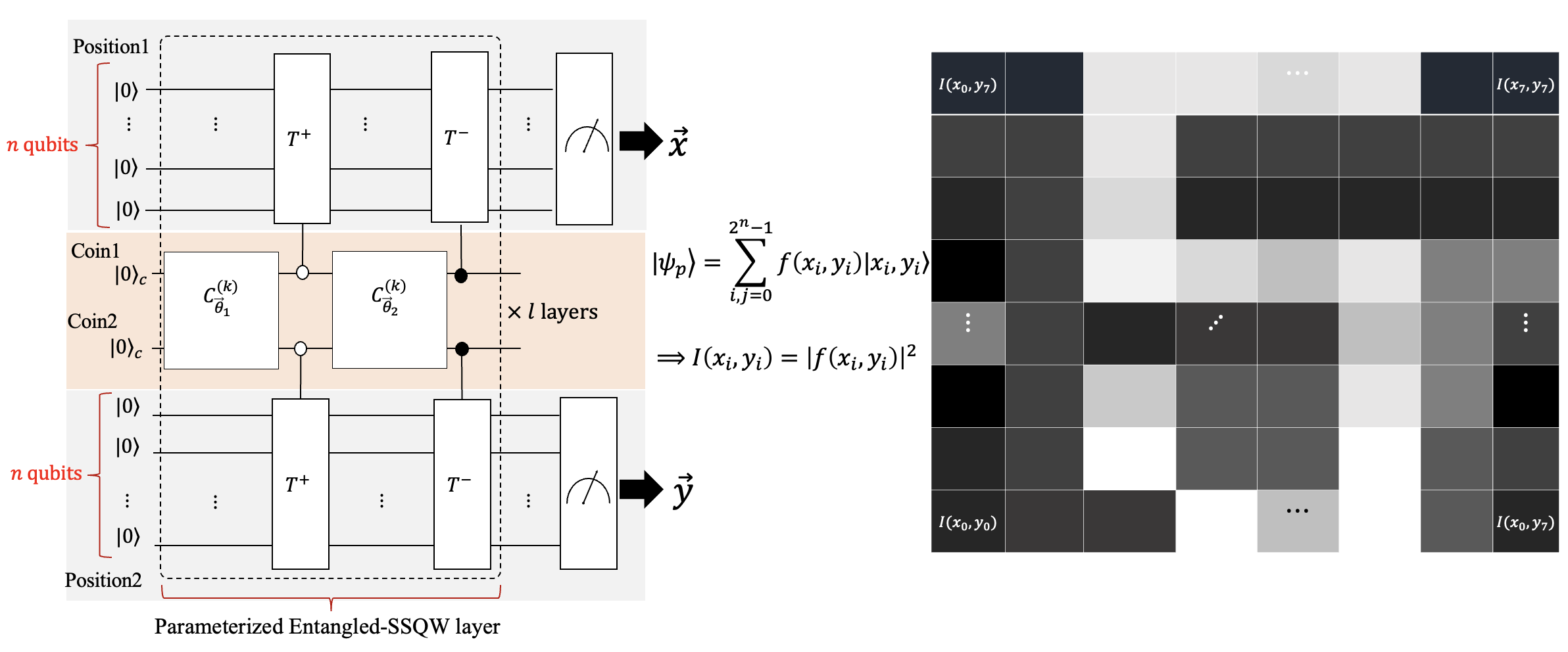}
    \caption{
        Conceptual diagram illustrating two entangled quantum walkers. Each walker occupies one spatial dimension (e.g., \(x\) and \(y\)), and both walkers share a joint coin Hilbert space of dimension \(4\). The entangled coin operator acts on the two-qubit coin state, inducing correlated movements in the 2D position space and enabling more intricate interference patterns.
    }
    \label{fig:entangled_QWs}
\end{figure}

To encode a two-dimensional figure into a quantum state, we represent the 2D state space as
\begin{equation}
|\psi\rangle = \sum_{x=0}^{2^n - 1} \sum_{y=0}^{2^n - 1} f(x,y) \, |x,y\rangle,
\end{equation}
where \(|x,y\rangle = |x\rangle \otimes |y\rangle\) denotes a position in the 2D lattice and \(n\) is the number of qubits per spatial dimension. In this formulation, the function \(f(x,y)\) encodes the complex amplitude at each position, and the intensity of a classical 2D image at coordinate \((x,y)\) is given by
\[
I(x,y) = |f(x,y)|^2.
\]
Thus, the probability of measuring the state \(|x,y\rangle\) in the quantum system is directly proportional to the intensity of the corresponding pixel in the classical image. This amplitude encoding is critical for quantum state preparation, as it enables the direct mapping of classical data into the quantum domain. Various methods—including those based on variational solvers and matrix product state representations—have been developed for efficient amplitude encoding \cite{Parl2021QSP,Zhang2022QSP,Yuan2023QSP}, with foundational approaches outlined in \cite{Nielsen2000}.

When extending QWs to higher-dimensional spaces or multiple walkers, one may utilize separate coin registers or merge them into a single joint coin register, thereby allowing entanglement among coin operators. As depicted in Figure~\ref{fig:entangled_QWs}, this approach, two entangled quantum walkers, assigns a shared coin Hilbert space of dimension \(4\) (\(2 \times 2\)), spanned by two qubits:
\[
\mathcal{H}_{c}^{(2)} = \mathrm{span}\Bigl\{|0\rangle_{x}\otimes|0\rangle_{y},\; |0\rangle_{x}\otimes|1\rangle_{y},\; |1\rangle_{x}\otimes|0\rangle_{y},\; |1\rangle_{x}\otimes|1\rangle_{y}\Bigr\}.
\]
Here, each basis state (e.g., \(|c_1 c_2\rangle = |c_1\rangle_{x}\otimes |c_2\rangle_{y}\)) represents a combined coin configuration for the two walkers. The position Hilbert space \(\mathcal{H}_{p}^{(x,y)}\) is two-dimensional, allowing independent movement along the \(x\) and \(y\) axes.

The entangled two-dimensional quantum walk is governed by the evolution operator:
\begin{equation}
\label{eq:W2D}
\hat{W}_{\text{2D-ent}} = \hat{S}_{xy}^- \,\hat{C}_{\vec{\boldsymbol{\theta}_2}}\,\hat{S}_{xy}^+ \,\hat{C}_{\vec{\boldsymbol{\theta}_1}},
\end{equation}
where:
\begin{enumerate}
    \item \textbf{Entangled Coin Operator:} \(\hat{C}_{\boldsymbol{\theta}_i}\) (\(i\in\{1,2\}\)) is a \(4\times 4\) unitary operator acting on the joint coin space, capable of entangling the coin states of both walkers. Its parameter set \(\vec{\boldsymbol{\theta}_i}\) is typically larger than that of a single-qubit operator.
    \item \textbf{Shift Operators:} The conditional shift operators \(\hat{S}_{xy}^\pm\) act on the 2D position space based on the joint coin state. They are defined by
    \begin{equation}
    \label{eq:Sxy}
    \hat{S}_{xy}^\pm = \sum_{c_1,c_2 \in \{0,1\}} \Bigl(|c_1c_2\rangle\langle c_1c_2|\Bigr) \otimes \hat{T}_{c_1,c_2}^\pm,
    \end{equation}
    where each \(\hat{T}_{c_1,c_2}^\pm\) is a conditional translation operator that shifts the walker’s position according to the specific coin state (e.g., \(\hat{T}_{00}^+\) may map \((x,y)\) to \((x+1,y)\), while \(\hat{T}_{01}^+\) may shift it to \((x,y+1)\)).
\end{enumerate}

This entangled configuration naturally extends our framework to generate two-dimensional distributions. For example, by adaptively tuning the joint coin parameters, our method can reproduce intricate spatial patterns such as MNIST digit images or other complex figures on an \(8 \times 8\) grid. Recent work on quantum-inspired image reconstruction \cite{ChangImage2024} further underscores the potential of such approaches in practical applications.

Overall, by leveraging the entanglement between coin registers, our Quantum Walk–Based Adaptive Distribution Generator is capable of capturing the intricate spatial correlations necessary for two-dimensional distribution generation.

\subsection{Variational Quantum Circuits}

Variational quantum circuits (VQCs) are parameterized quantum circuits whose gate parameters are optimized via a classical feedback loop to perform specific computational tasks, such as approximating target probability distributions \cite{Benedetti19,Cerezo21}. The typical training process involves the following steps:
\begin{enumerate}
    \item \textbf{Initialization:} Construct a quantum circuit with an initial parameter set \(\vec{\theta}\).
    \item \textbf{Quantum Processing:} Evolve an input state through the circuit to produce the output state \(|\psi(\vec{\theta})\rangle\).
    \item \textbf{Measurement:} Measure the output state to obtain a simulated probability distribution \(P_{\text{sim}}(x)\).
    \item \textbf{Cost Evaluation:} Define a cost function \(E(\vec{\theta})\) (e.g., mean-square error or Kullback-Leibler divergence) that quantifies the discrepancy between \(P_{\text{sim}}(x)\) and a target distribution \(P_{\text{target}}(x)\).
    \item \textbf{Parameter Update:} Adjust the circuit parameters using a classical optimization algorithm, typically following the update rule:
    \begin{equation}
        \vec{\theta}_{\text{new}} = \vec{\theta}_{\text{old}} - \eta \nabla_{\vec{\theta}} E(\vec{\theta}),
    \end{equation}
    where \(\eta\) is the learning rate.
\end{enumerate}

This hybrid quantum-classical optimization loop enables VQCs to efficiently learn and reproduce complex, high-dimensional distributions, making them a cornerstone for quantum generative models.

\subsection{Integration with Variational QW-Based Quantum Circuits on CUDA-Q}

Our integrated framework synergistically combines the adaptive capabilities of VQCs with the dynamical evolution of QWs to form a QWs-based ADG. In this approach, the coin operators within the SSQWs are implemented as parameterized U3 gates (see Equation \ref{eq:Coinoperator}). The QWs-based ADG continuously updates these coin parameters via a classical gradient descent procedure to minimize the discrepancy between the simulated probability distribution \(P_{\text{sim}}(x)\) and the target distribution \(P_{\text{target}}(x)\):
\begin{equation}
\begin{pmatrix}
\theta_{\text{new}} \\
\phi_{\text{new}} \\
\lambda_{\text{new}}
\end{pmatrix}
=
\begin{pmatrix}
\theta_{\text{old}} \\
\phi_{\text{old}} \\
\lambda_{\text{old}}
\end{pmatrix}
-
\begin{pmatrix}
\eta_\theta \, \dfrac{\partial E}{\partial \theta} \\
\eta_\phi \, \dfrac{\partial E}{\partial \phi} \\
\eta_\lambda \, \dfrac{\partial E}{\partial \lambda}
\end{pmatrix}.
\end{equation}

To handle the computational challenges associated with simulating large state spaces and optimizing numerous parameters, our hybrid framework is implemented on the CUDA-Q platform \cite{Kim23}. CUDA-Q is an open-source quantum computing platform that enables seamless hybrid programming across quantum and classical systems within a unified framework. In this study, we leverage CUDA-Q to simulate quantum circuits, powered by its underlying cuQuantum library\cite{10313722} with GPU acceleration, substantially reducing computation time and enhancing scalability. This integration enables our QWs-based ADG to efficiently learn and reproduce complex probability distributions in both one-dimensional and two-dimensional settings.

By entangling the coin registers of two quantum walkers, the movements along the \(x\) and \(y\) axes become correlated, which leads to richer interference patterns and enables the generation of more complex spatial distributions \cite{Chandrashekar10,Mallick16}. This entangled configuration naturally extends our framework to address two-dimensional distribution generation, as demonstrated by our ability to generate digit patterns on an \(8 \times 8\) grid.

Overall, by leveraging the flexibility of DTQWs, SSQWs, and their entangled extensions, our framework provides a robust foundation for controlled quantum simulation and serves as a critical component of our Quantum Walk–Based ADG.

\section{Results and Discussion}
\label{sec:results}

In this section, we present the performance of our QWs-based ADG framework across a diverse set of target distributions. We begin by demonstrating its efficacy in reproducing six distinct one-dimensional (1D) distributions: (1) the NVDA return distribution, (2) beta, (3) binomial, (4) bimodal, (5) exponential, and (6) poisson. We then apply our approach to simulate a log-normal distribution for European call option pricing, highlighting its relevance to financial applications. Finally, we illustrate the expressiveness of entangled quantum walkers by generating two-dimensional (2D) patterns representing digit images.

\subsection{One-Dimensional Distribution Modeling}

\begin{figure}
    \centering
    % Example placeholders; replace with actual figures for each distribution
    \includegraphics[width=0.32\linewidth]{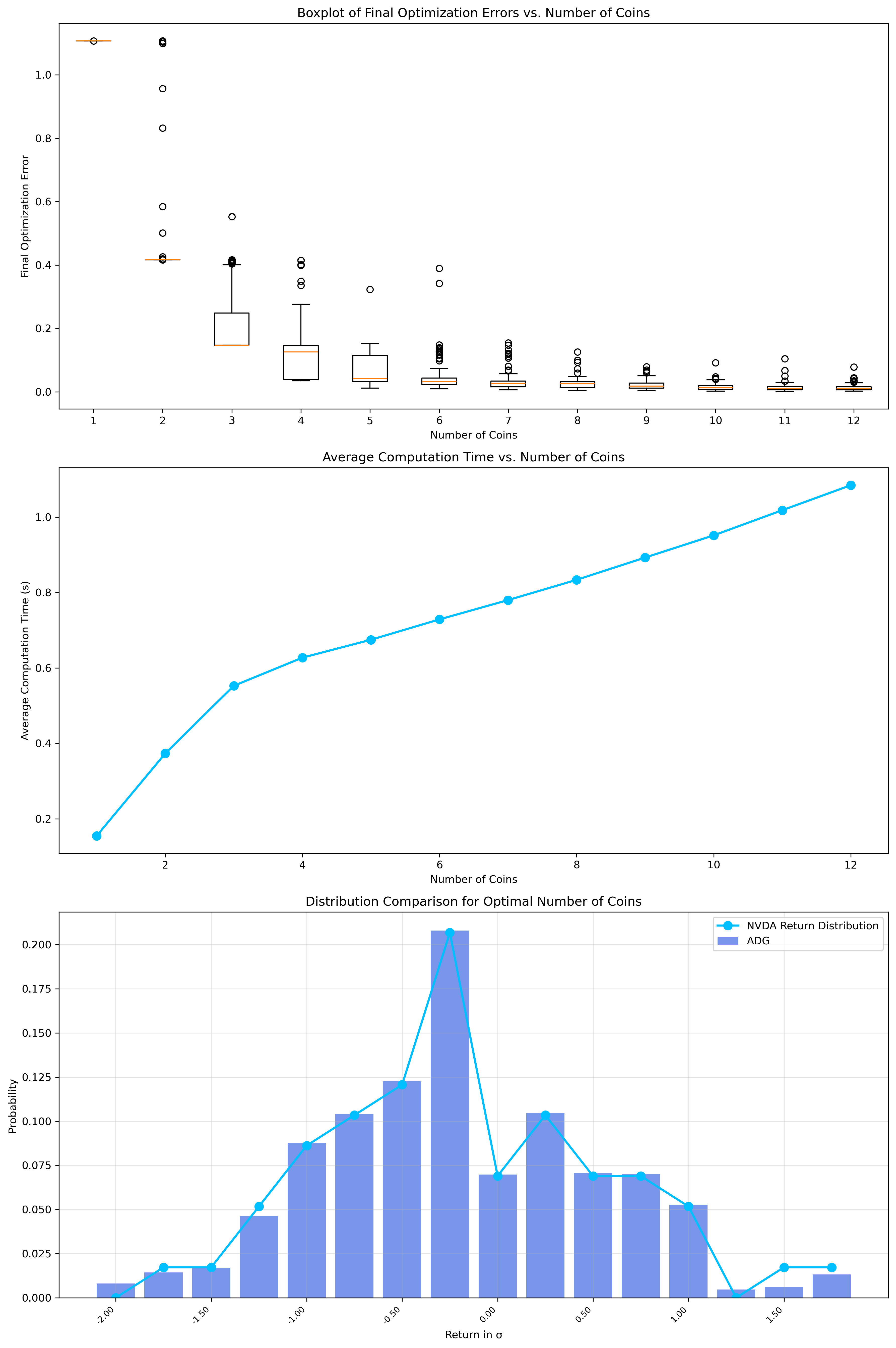}
    \includegraphics[width=0.32\linewidth]{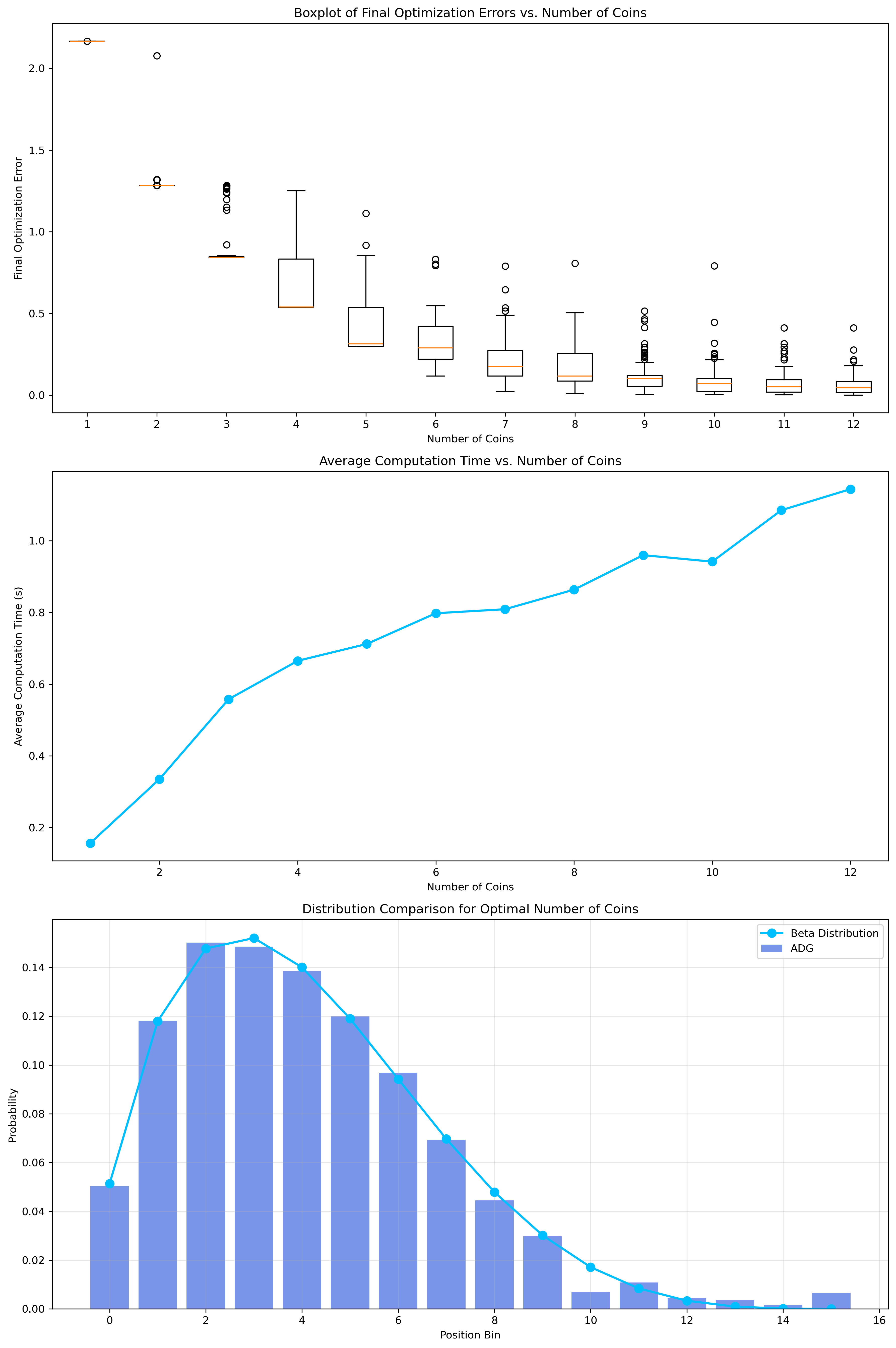}
    \includegraphics[width=0.32\linewidth]{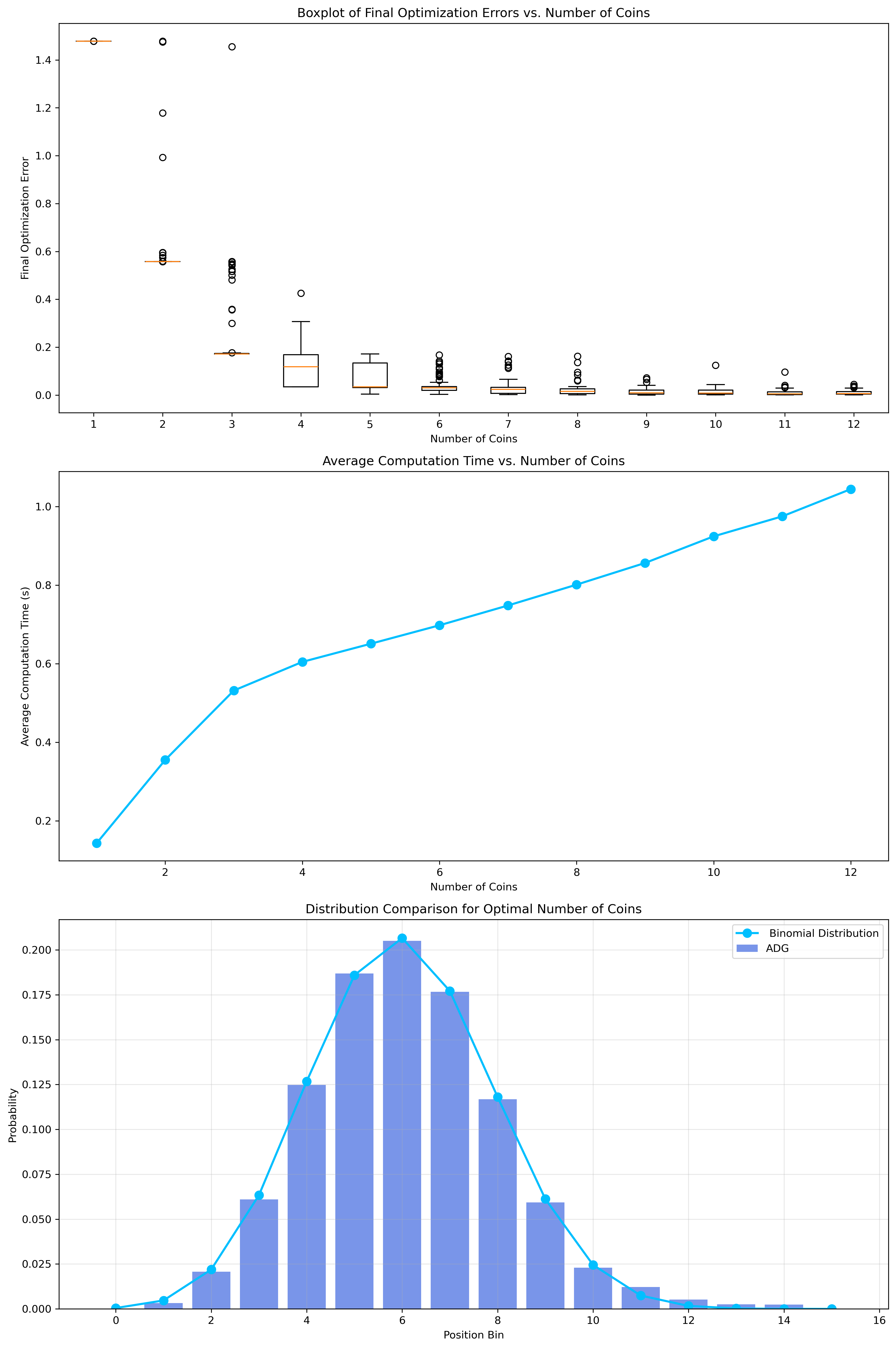}
    \includegraphics[width=0.32\linewidth]{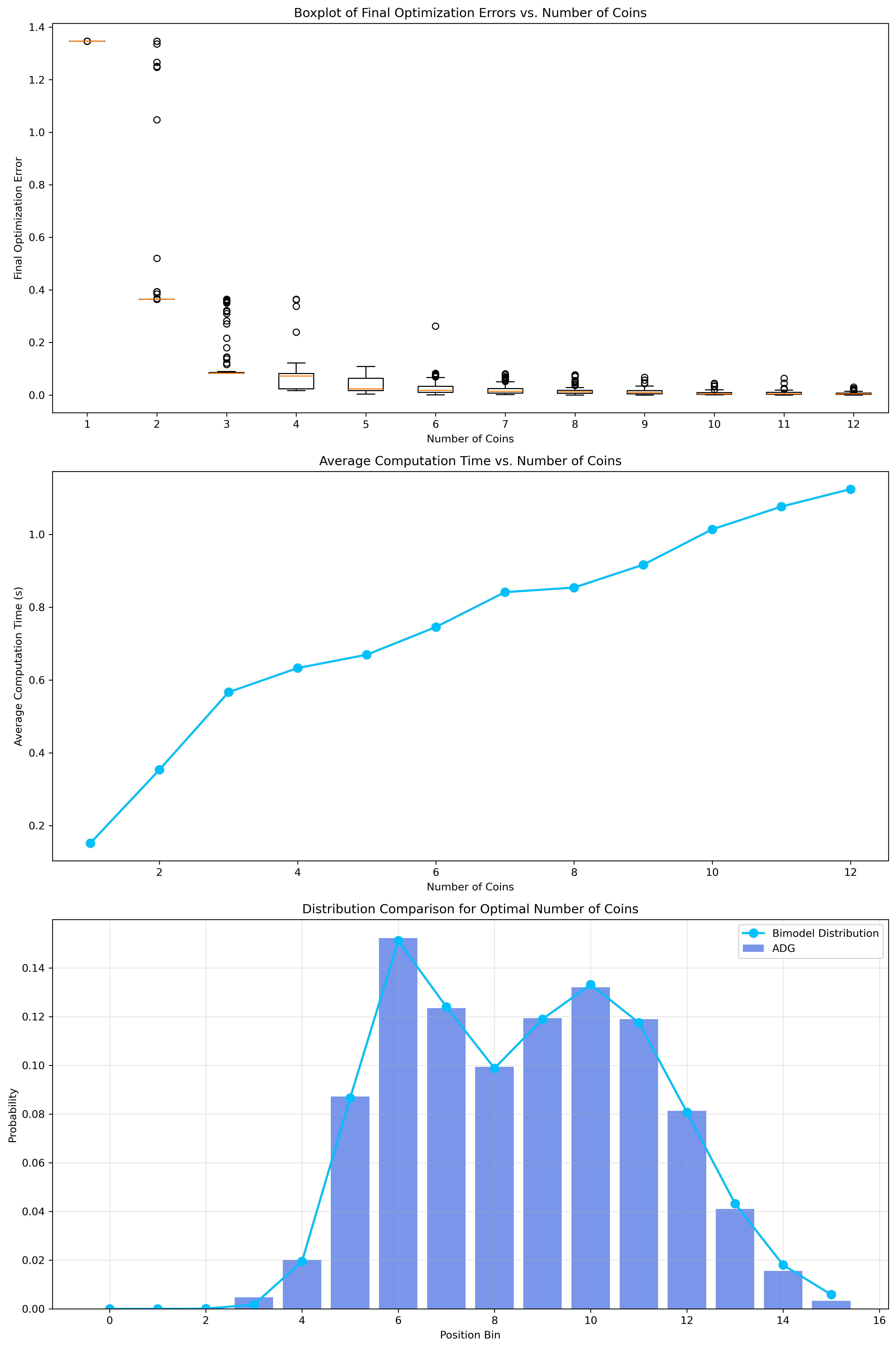}
    \includegraphics[width=0.32\linewidth]{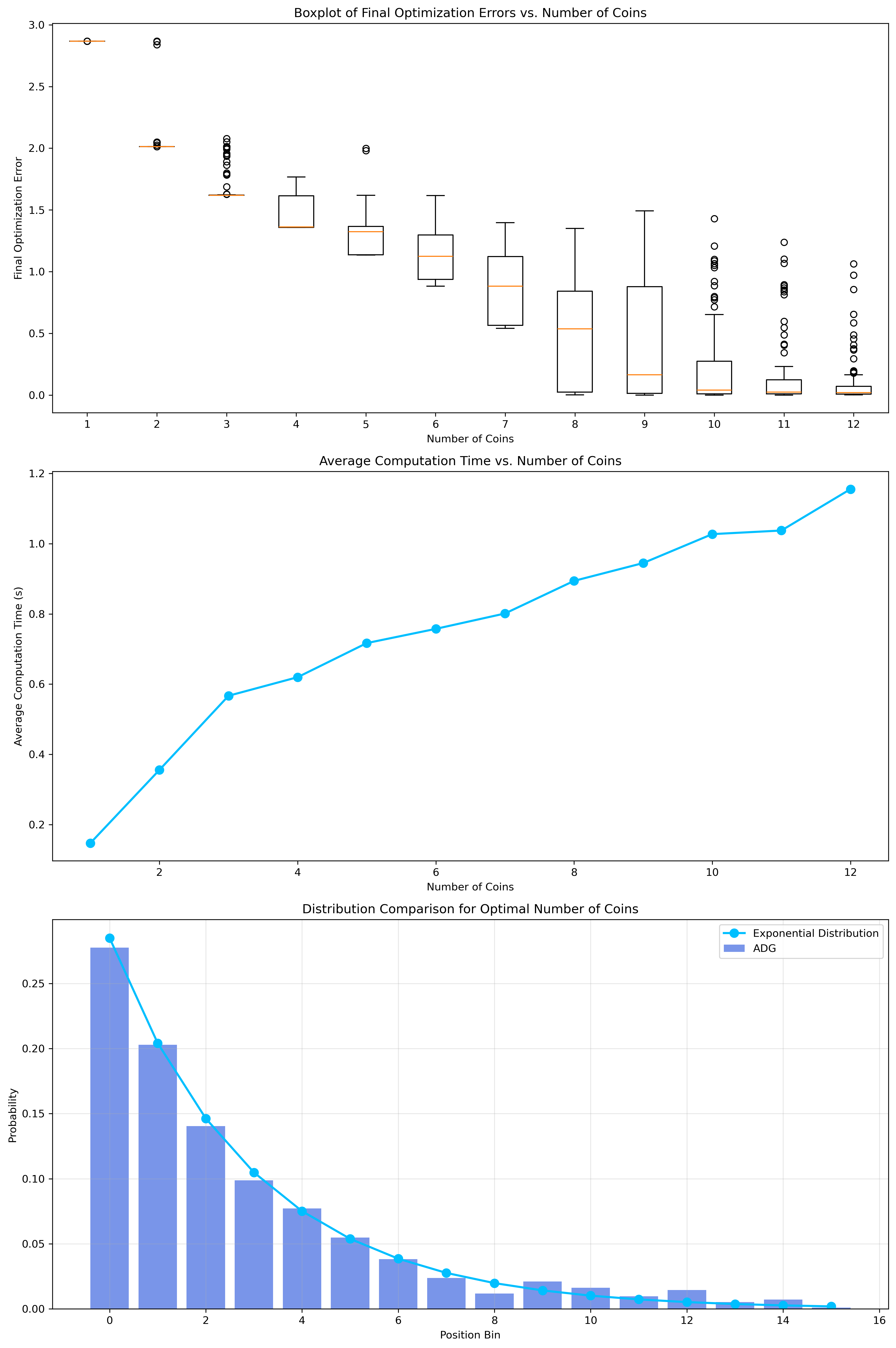}
    \includegraphics[width=0.32\linewidth]{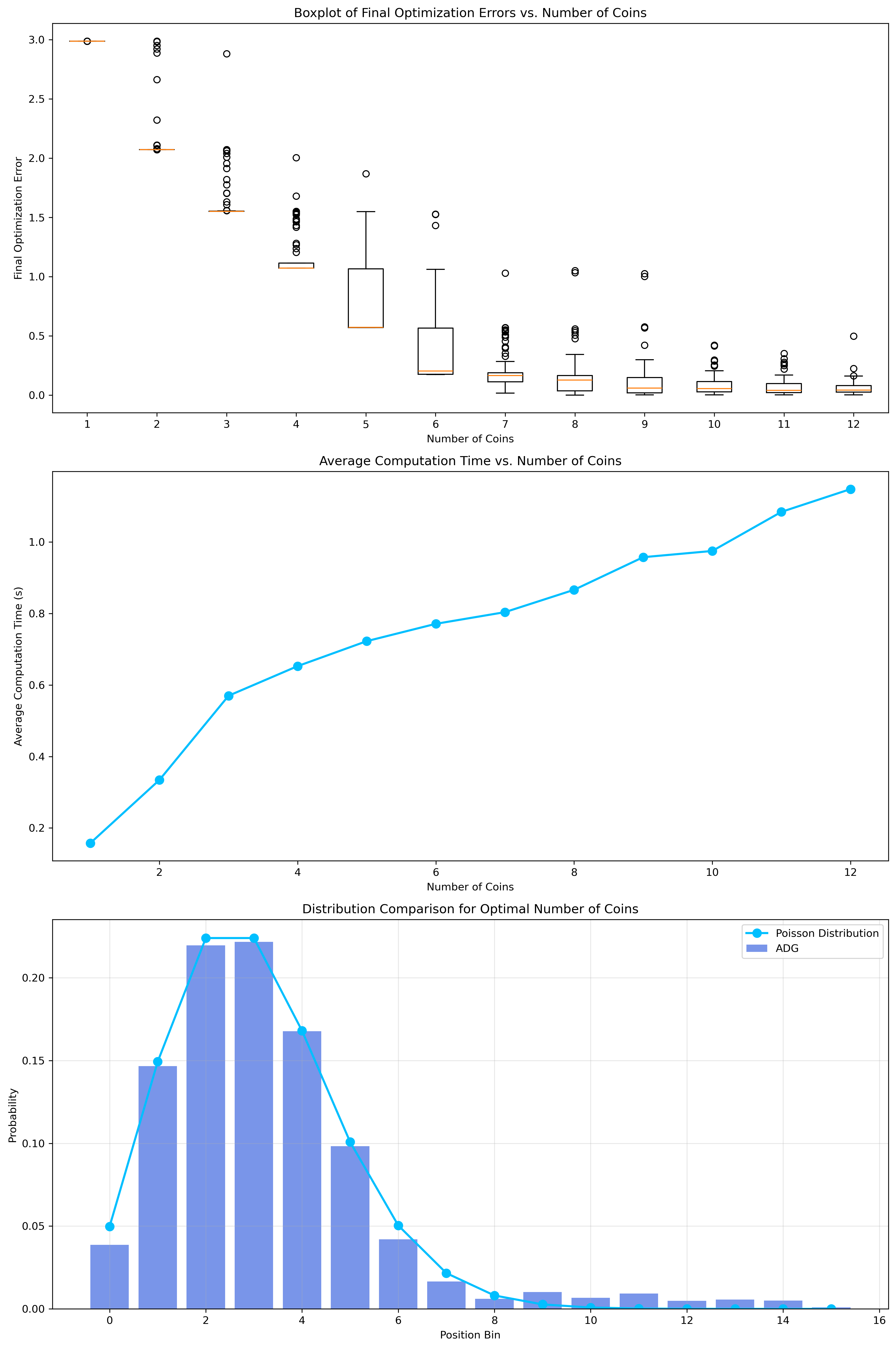}
    \caption{Representative results for the six one-dimensional target distributions: NVDA returns, beta, binomial, bimodal, exponential, and poisson. Each set of plots shows (top) a boxplot of final optimization errors, (middle) average computation time, and (bottom) a comparison between the QWs-based ADG-simulated distribution and the target distribution.}
    \label{fig:results_1D}
\end{figure}

Figure~\ref{fig:results_1D} provides examples of our results for each of the six 1D distributions. Each subfigure is divided into three panels:
\begin{enumerate}
    \item \textbf{Boxplot of Final Optimization Errors vs. Number of Coins:} We vary the number of coins (representing the number of independent coin qubits or coin steps) and observe how the final optimization error decreases as the number of coins increases. Generally, a larger number of coins yields greater expressiveness and lower final errors, though at the cost of increased computational time.
    \item \textbf{Average Computation Time vs. Number of Coins:} The second panel shows how the average computation time scales with the number of coins. Although more coins improve accuracy, they also increase computational overhead.
    \item \textbf{Distribution Comparison for the Optimal Number of Coins:} The final panel compares the QWs-based ADG-simulated distribution (blue bars) with the target distribution (solid line or markers) at the optimal coin configuration. These results confirm that our approach closely matches the desired target across different distribution shapes.
\end{enumerate}

Across all these distributions, the QWs-based ADG mechanism successfully adapts the coin parameters to minimize the discrepancy between the quantum walk’s simulated distribution and the target. Notably, distributions with heavier tails (e.g., certain financial returns) may require a larger number of coins to capture subtle features. Overall, our results demonstrate that the QWs-based ADG is versatile enough to model both  real-world data such as NVDA stock returns and well-known statistical distributions (beta, binomial, bimodal, exponential, and Poisson).

\subsection{Log‑Normal Distribution for Option Pricing and Error Analysis}

Log‑normal distributions are fundamental in financial modeling, underpinning the Black–Scholes framework by modeling asset prices as a geometric Brownian motion \cite{BlackScholes73}. Prior studies in quantum option pricing have leveraged quantum amplitude amplification and estimation to achieve quadratic reductions in sampling error \cite{Brassard02, Rebentrost18,Stamatopoulos20}.

\begin{figure}[ht]
    \centering
    \includegraphics[width=0.5\linewidth]{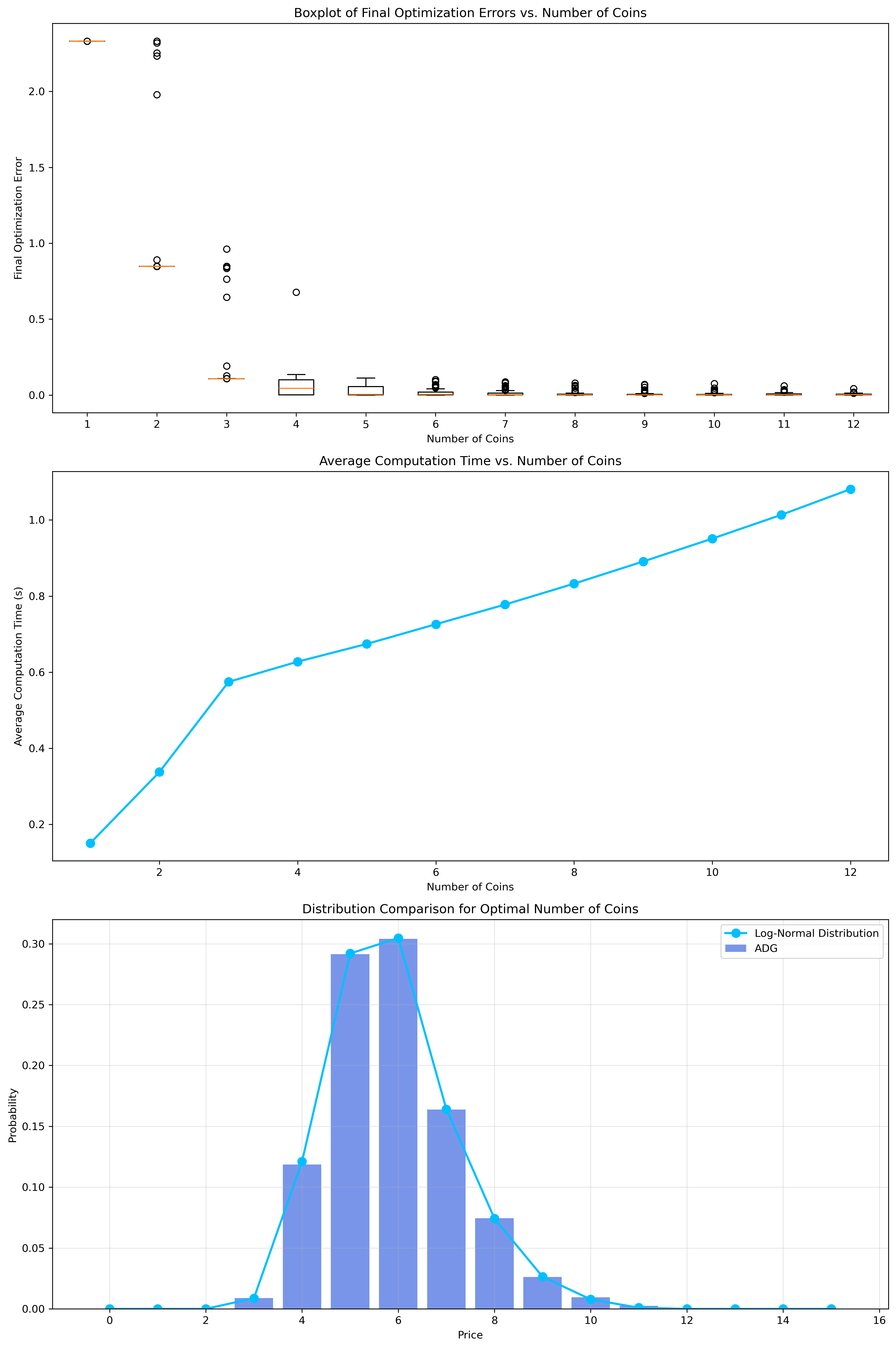}
    \caption{CUDA‑Q–simulated log‑normal distribution from our QWs‑based ADG, used for pricing a European call option with \(S=6.0\), \(\sigma=0.4\), \(r=0.04\), and \(T=90/365\).}
    \label{fig:lognormal_distribution}
\end{figure}

In our QWs‑based ADG framework—implemented entirely via high‑performance simulation on the CUDA‑Q platform—we allocate five qubits (one for the coin register and four for the position register), yielding a discrete price grid of \(2^4 = 16\) points. We then optimize the coin parameters so that, after \(t\) evolution steps, the resulting position‑space probability distribution \(P(x)\) closely approximates the target log‑normal density. Figure~\ref{fig:lognormal_distribution} shows the calibrated discrete distribution for a spot price \(S=6.0\), volatility \(\sigma=0.4\), risk‑free rate \(r=0.04\), and time to maturity \(T=90/365\).

From this distribution, we compute the expected payoff of a European call option by
\[
\mathbb{E}\bigl[\max(S_T - K,0)\bigr]
\;\approx\;
\sum_{i=1}^{16} P_{\mathrm{sim}}(S_i)\,\max(S_i - K,0)\,e^{-rT}\,,
\]
Moreover, compare these simulated prices against the analytical Black–Scholes values in Table~\ref{tab:option_pricing}. The QWs‑based ADG closely tracks the benchmark for at‑ and in‑the‑money strikes, though errors increase significantly for deep out‑of‑the‑money options.

\begin{table}[htbp]
    \centering
    \caption{European call prices: Black–Scholes vs.\ CUDA‑Q–simulated QWs‑ADG}
    \label{tab:option_pricing}
    \begin{tabular}{|c|c|c|c|}
        \hline
        \(K\) & Black–Scholes & QWs‑ADG (CUDA‑Q) & Error (\%) \\
        \hline
         1 & 5.0098 & 4.8389 &  3.41  \\
         2 & 4.0196 & 3.8487 &  4.25  \\
         3 & 3.0295 & 2.8588 &  5.63  \\
         4 & 2.0457 & 1.8778 &  8.21  \\
         5 & 1.1447 & 1.0141 & 11.41  \\
         6 & 0.5024 & 0.4390 & 12.60  \\
         7 & 0.1745 & 0.1650 &  5.41  \\
         8 & 0.0502 & 0.0531 &  5.92  \\
         9 & 0.0126 & 0.0151 & 19.87  \\
        10 & 0.0029 & 0.0030 &  4.43  \\
        \hline
    \end{tabular}
\end{table}

\paragraph{Error Analysis}
The observed discrepancies arise from two main sources:
\begin{enumerate}
    \item \textbf{Discretization Error:} Our five‑qubit setup defines a finite grid of 16 price points, while the Black-Scholes model assumes a continuous log-normal distribution. This discretization introduces the most significant errors in the distribution tails, where the probability mass is sparse.
    \item \textbf{Sampling and Optimization Noise:} The iterative ADG parameter updates and finite sampling in the simulated measurements introduce variance, particularly affecting payoffs sensitive to low‑probability events.
\end{enumerate}

\paragraph{Discussion}
Despite these limitations, our GPU-accelerated simulations confirm that SSQWs can capture the essential features of log-normal distributions for option pricing. Improving precision will require (i) increasing the number of position qubits to refine grid resolution, (ii) employing advanced error‑mitigation and denoising techniques, and (iii) ultimately validating the approach on physical quantum hardware. This work lays the groundwork for practical quantum‑enhanced financial simulations, bridging high‑performance emulation with future real‑device implementations.

\subsection{Entangled Quantum Walks for Two-Dimensional Pattern Generation}

\begin{figure*}[htb]
    \centering
    %
    % Row 1
    \subfloat[Digit 0]{
        \includegraphics[width=0.18\linewidth]{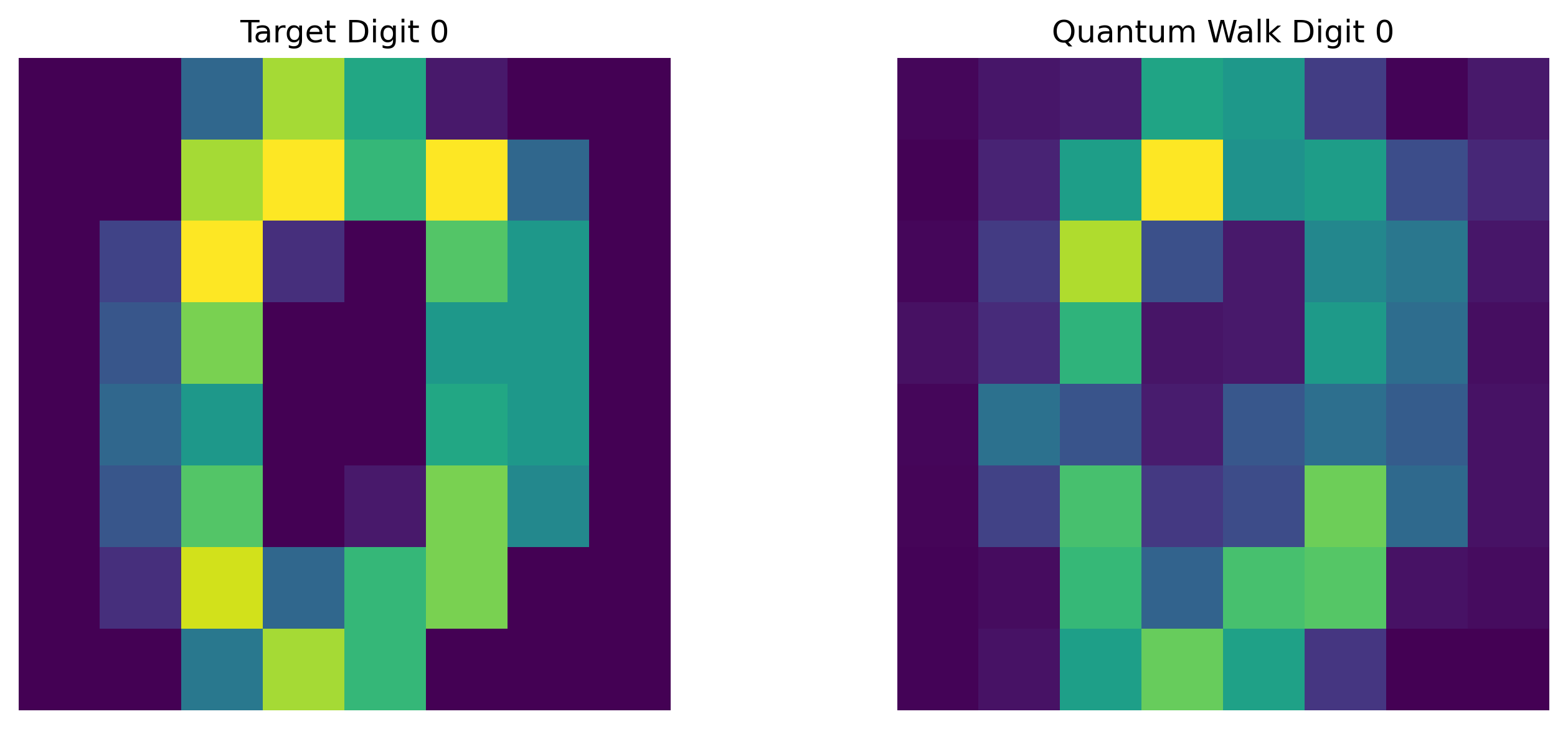}
    }
    \subfloat[Digit 1]{
        \includegraphics[width=0.18\linewidth]{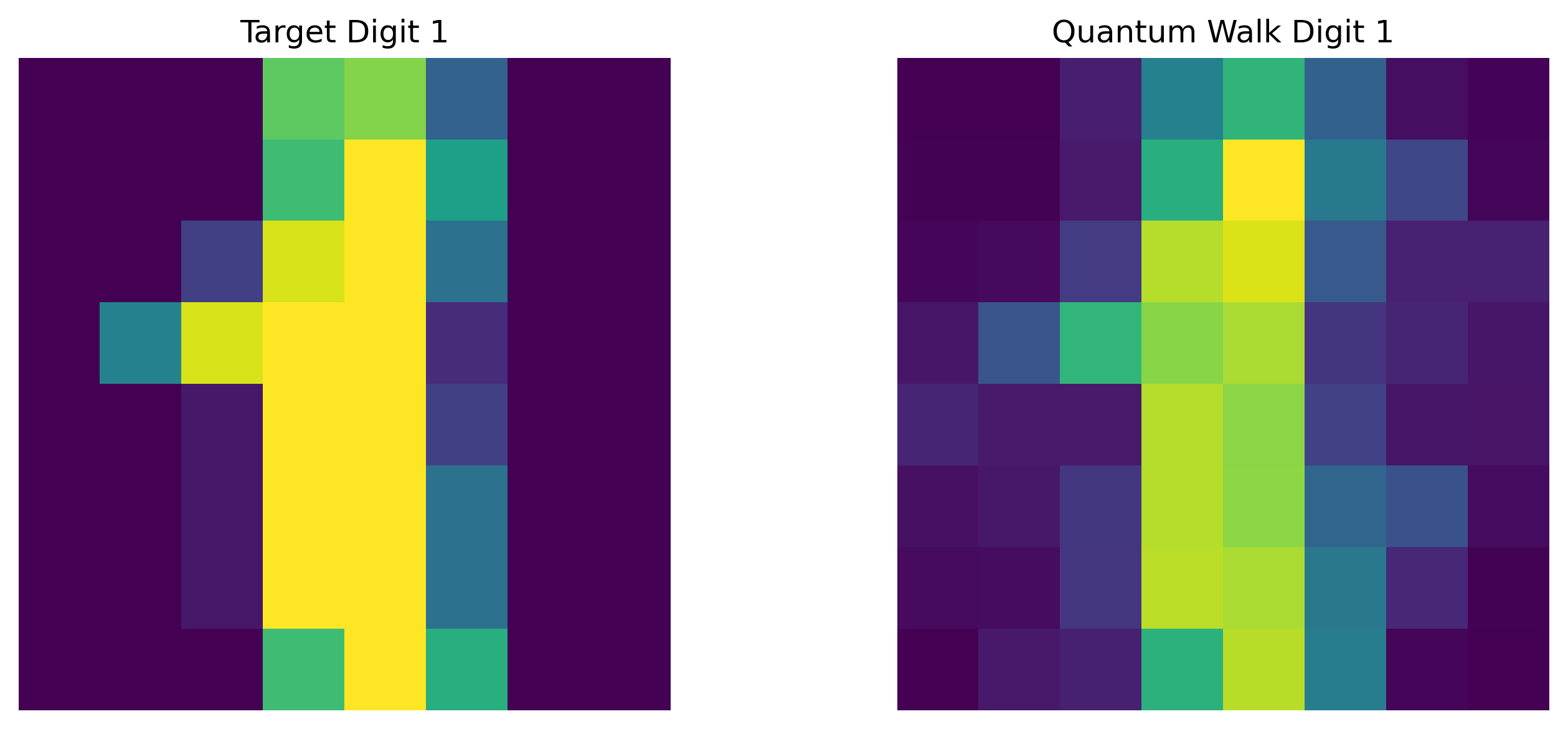}
    }
    \subfloat[Digit 2]{
        \includegraphics[width=0.18\linewidth]{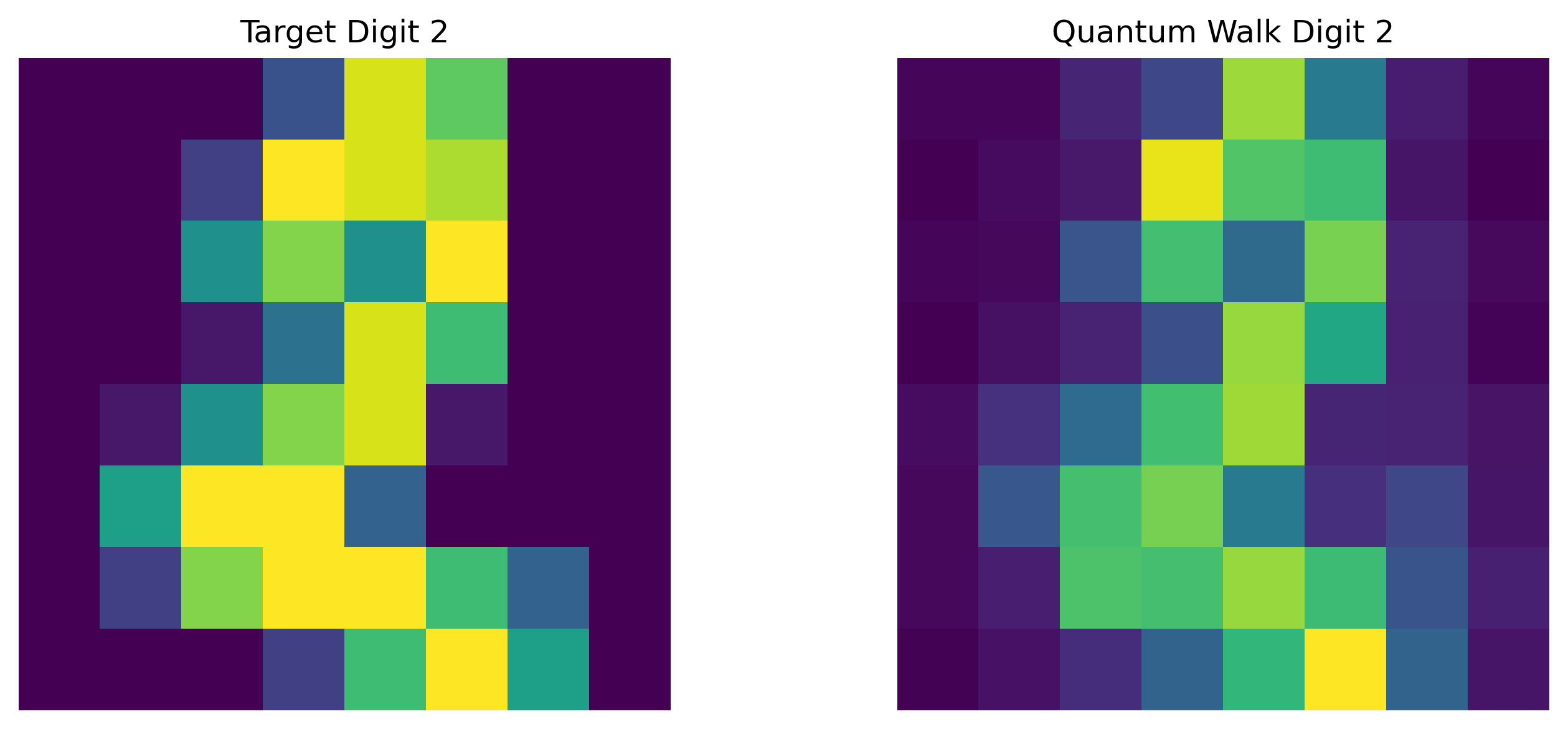}
    }
    \subfloat[Digit 3]{
        \includegraphics[width=0.18\linewidth]{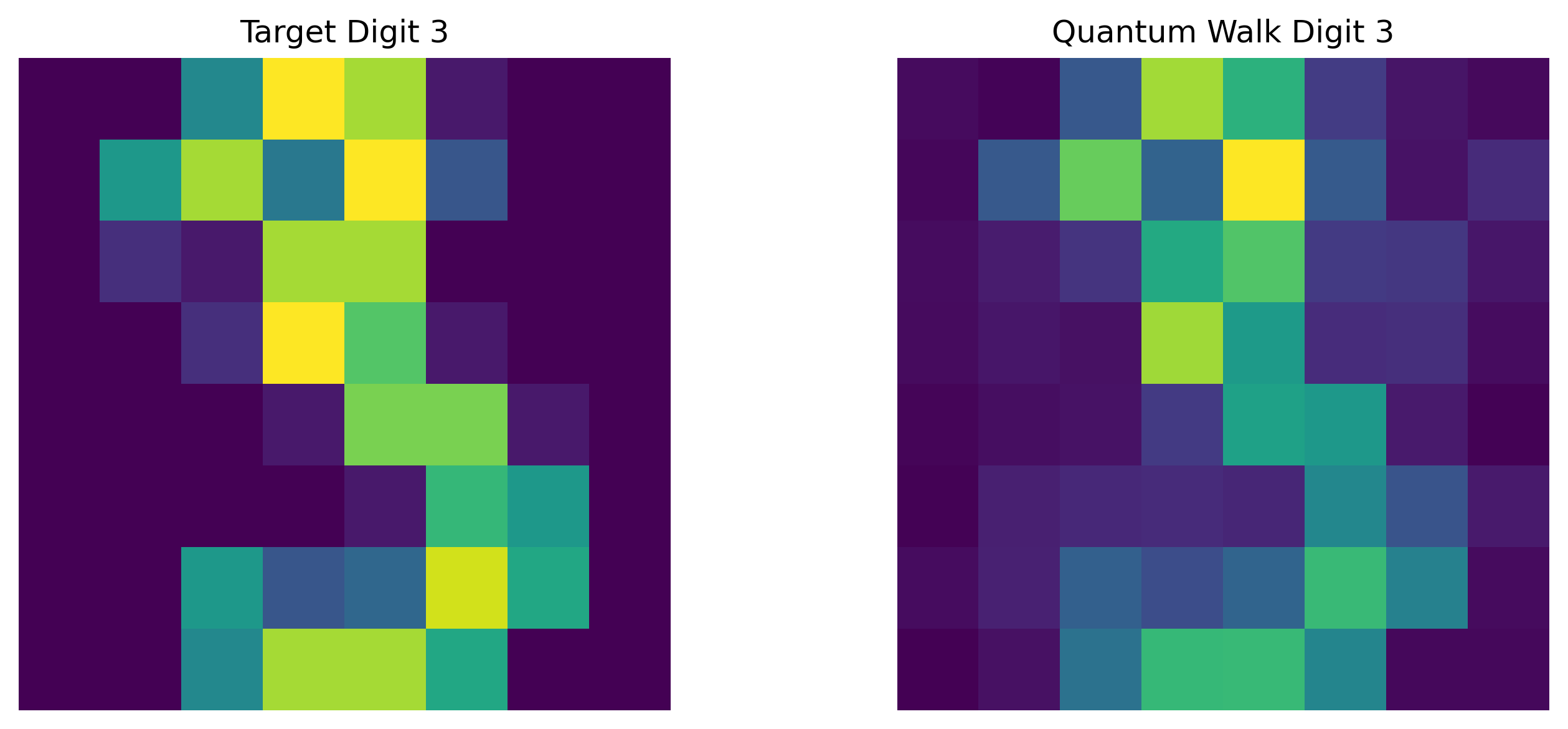}
    }
    \subfloat[Digit 4]{
        \includegraphics[width=0.18\linewidth]{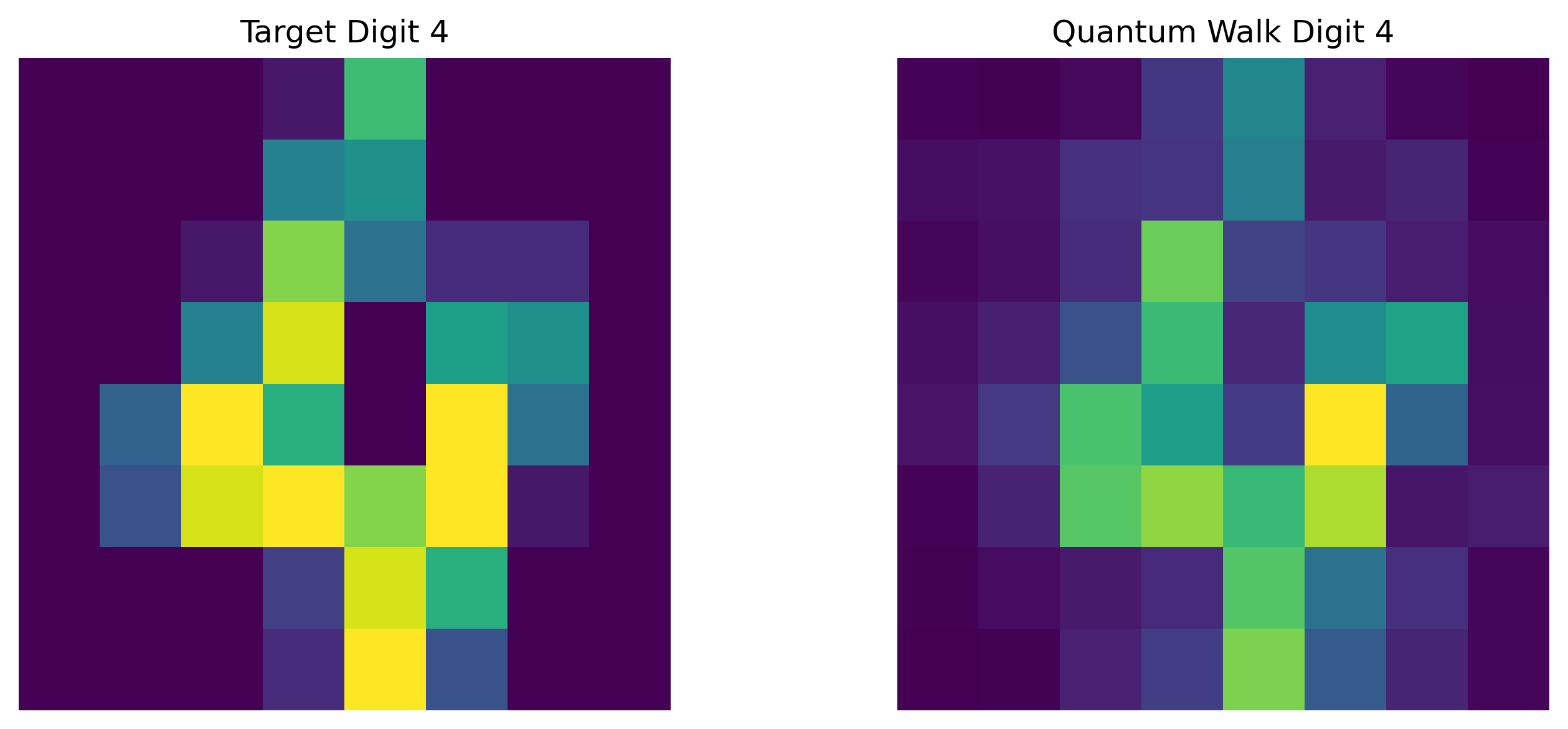}
    }
    \\
    % Row 2
    \subfloat[Digit 5]{
        \includegraphics[width=0.18\linewidth]{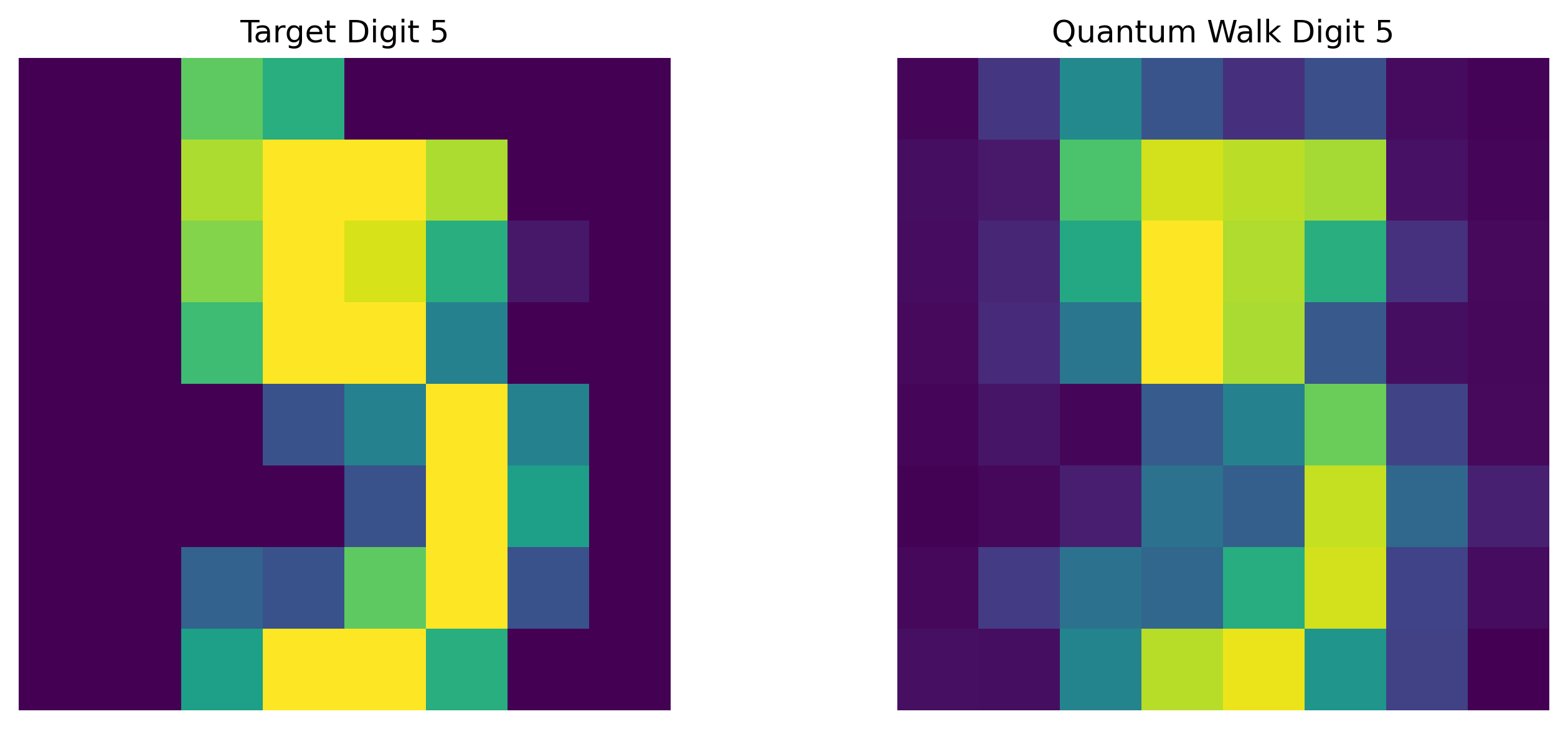}
    }
    \subfloat[Digit 6]{
        \includegraphics[width=0.18\linewidth]{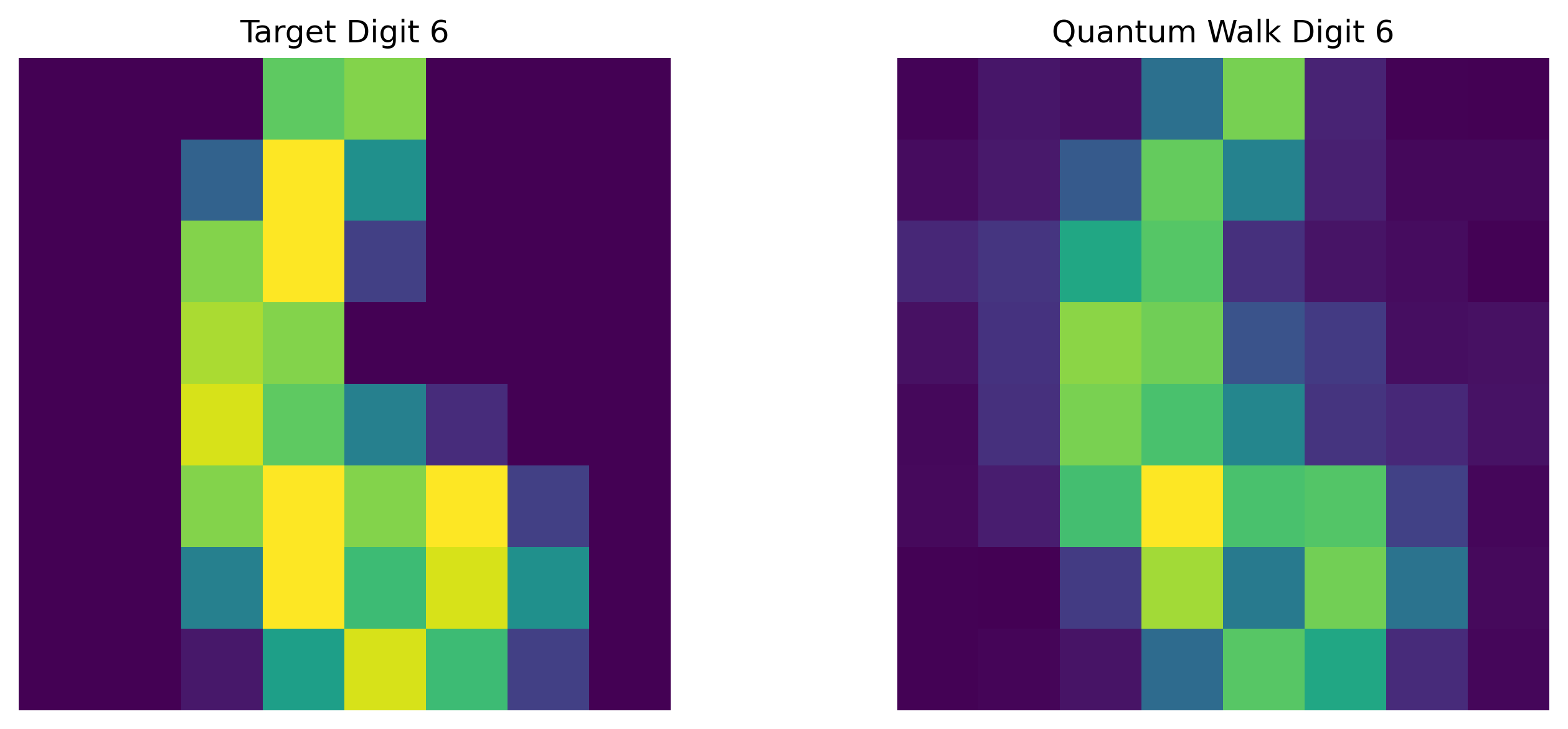}
    }
    \subfloat[Digit 7]{
        \includegraphics[width=0.18\linewidth]{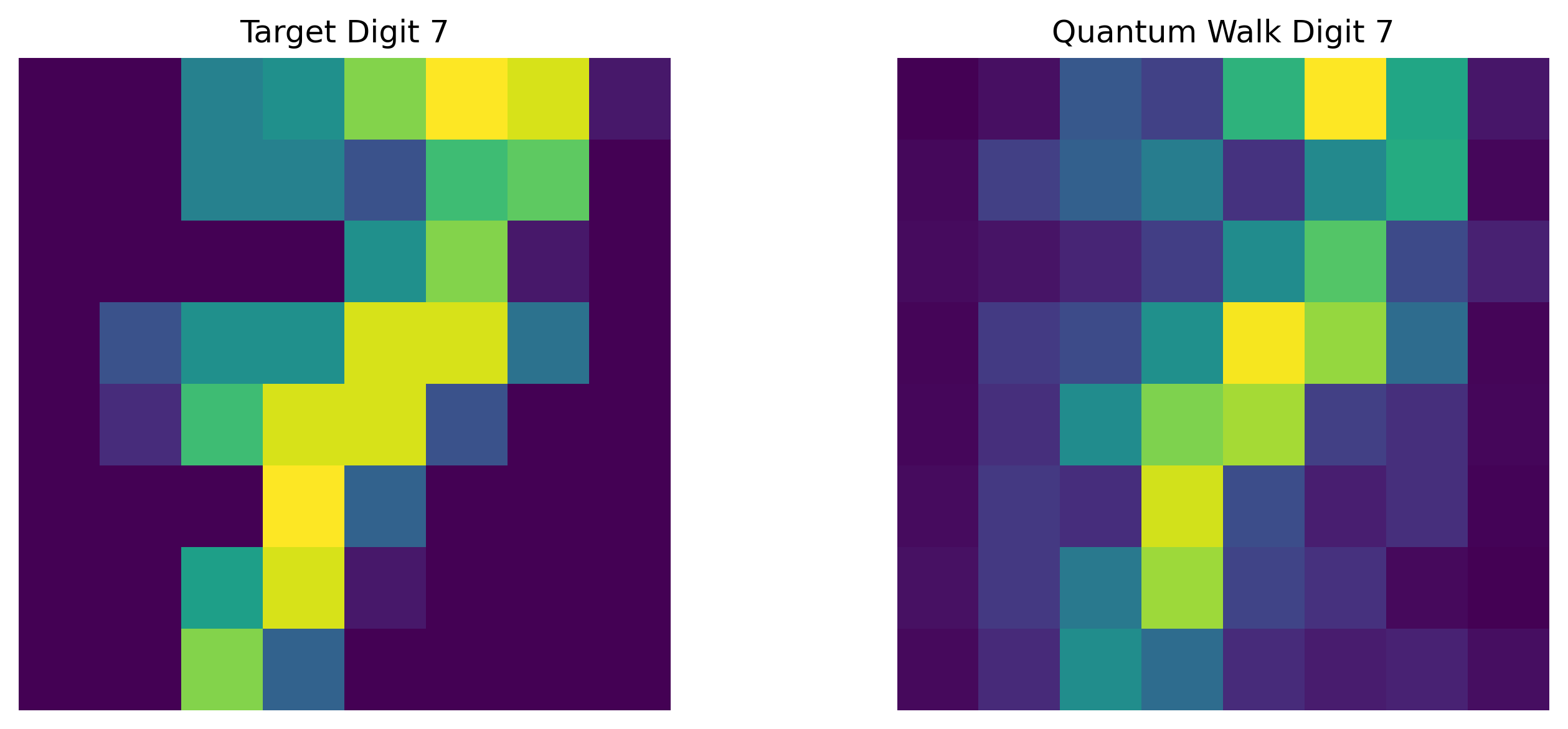}
    }
    \subfloat[Digit 8]{
        \includegraphics[width=0.18\linewidth]{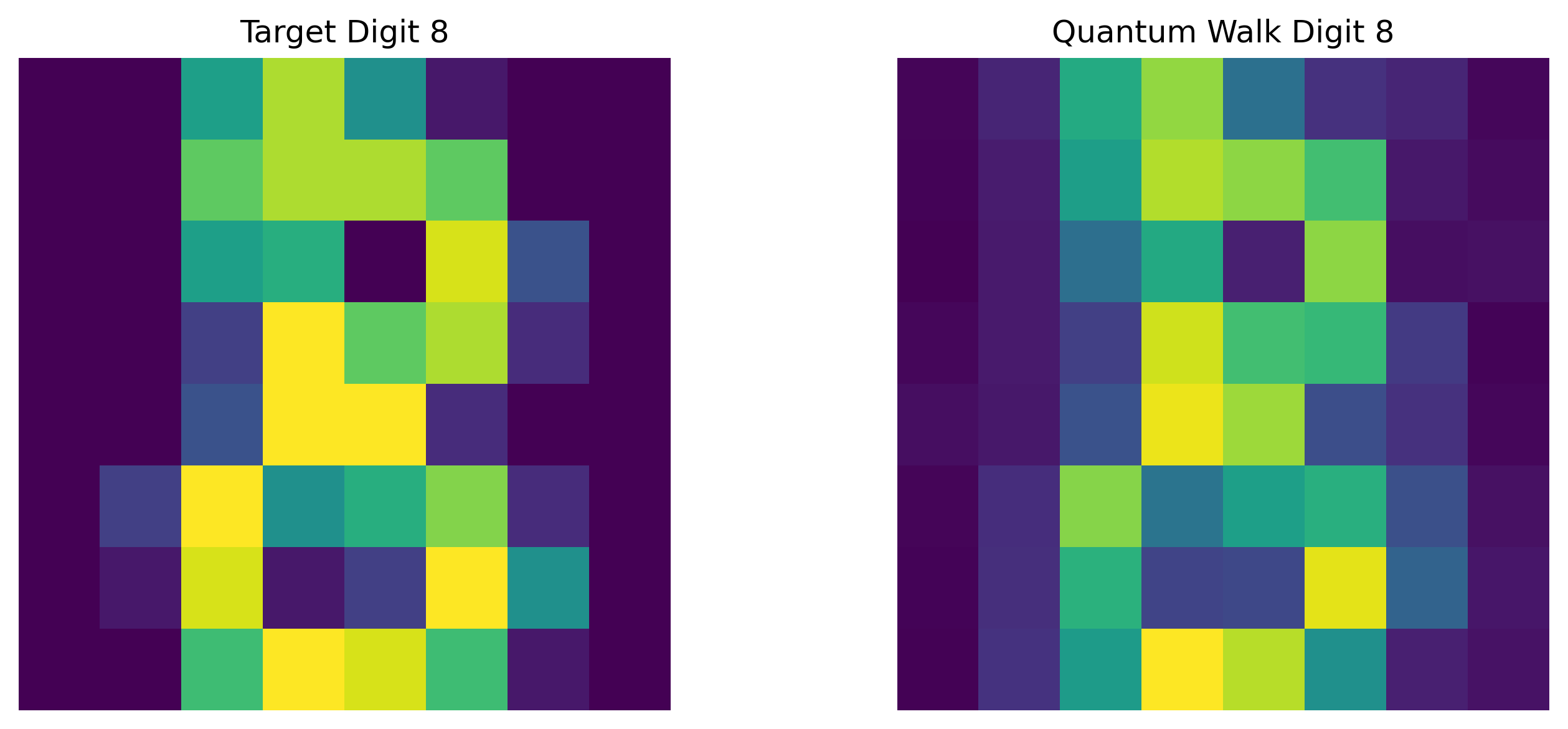}
    }
    \subfloat[Digit 9]{
        \includegraphics[width=0.18\linewidth]{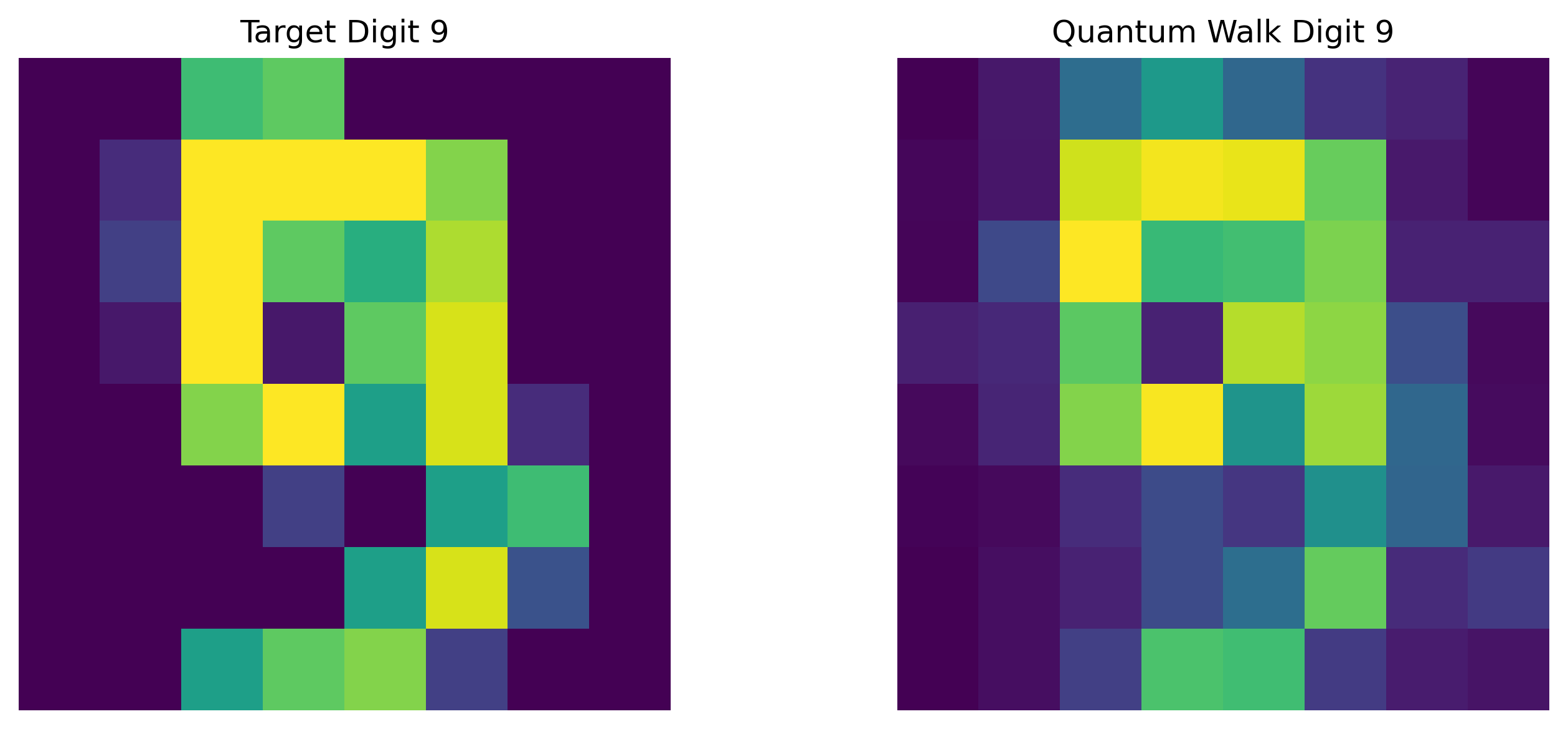}
    }
    \caption{
        Examples of 2D digit generation (0--9) using entangled quantum walkers.
        Each subfigure shows the final QWs-based ADG-simulated distribution for a specific digit on an 8\(\times\)8 grid.
        The coin spaces of two quantum walkers are entangled to capture the spatial structure of each digit.
    }
    \label{fig:entangled_digits}
\end{figure*}

Finally, we showcase the ability of our QWs-based ADG to generate two-dimensional patterns by entangling the coin spaces of two independent quantum walkers. Figure~\ref{fig:entangled_digits} compares the target 8\(\times\)8 digit patterns (left) with the simulated distributions (right) for several digits (0--9). The entanglement between coin registers allows interference effects to propagate across both walkers, enabling the model to capture complex spatial features. The QWs-based ADG mechanism is extended to jointly optimize the parameters of both coin registers, ensuring convergence to the desired 2D patterns.

This experiment underscores the versatility of our QWs-based ADG framework, extending beyond one-dimensional probability modeling to complex 2D generative tasks. By exploiting local unitary evolution and entanglement, the model can capture intricate correlations and spatial structures, making it a promising tool for quantum machine learning, image processing, and other high-dimensional generative problems.

\section{Conclusion and Future Work}
\label{sec:conclusion}

This work presents a QW-based ADG that combines variational quantum circuits with split-step and entangled quantum walks to learn and reproduce complex probability distributions. Our QWs‑based ADG successfully captured a variety of one‑dimensional distributions—such as binomial, bimodal, exponential, Poisson, and empirical financial returns (e.g., NVDA)—and accurately simulated a log‑normal density for European call option pricing. Extending this framework to two dimensions via entangled coin registers, we demonstrated the generation of structured patterns (MNIST‑style digits) on an \(8\times8\) grid with high fidelity.

Key strengths of our approach include:
\begin{itemize}
    \item \textbf{Adaptive Coin Tuning:} Gradient‑based updates of the U3 coin parameters steer the walker’s evolution toward arbitrary target distributions while preserving quantum coherence and interference.
    \item \textbf{Modular Evolution:} The split-step protocol decomposes the dynamics into simple, local unitaries, simplifying optimization and preserving entanglement across steps.
    \item \textbf{Scalability via CUDA‑Q:} GPU‑accelerated simulation enables rapid prototyping on up to five qubits, laying the groundwork for future hardware deployment.
    \item \textbf{Higher‑Dimensional Generative Power:} Entangling multiple coin registers extends naturally to 2D and beyond, opening avenues in quantum machine learning and image synthesis.
\end{itemize}

\paragraph{Future Work}
Building on these results, we plan to explore:
\begin{itemize}
    \item \textbf{Analytic Characterization of Coin Dynamics:} Systematically map how variations in \(\theta\), \(\phi\), and \(\lambda\) translate into statistical features (e.g., mean, variance, skewness, kurtosis) and entanglement measures of the resulting distribution.
    \item \textbf{Multi‑Coin and Continuous‑Variable Extensions:} Generalize the ADG to multi‑coin setups and continuous‑variable quantum walks, enabling finer discretization and richer generative models.
    \item \textbf{Experimental Realization and Noise Mitigation:} Deploy the QWs‑ADG on near‑term quantum hardware, developing error‑mitigation and noise‑resilient variants to maintain distribution fidelity under realistic device limitations.
\end{itemize}

By following these directions, we anticipate advancing the theoretical understanding and practical implementation of quantum walk–based generative models, paving the way for quantum-accelerated applications in finance, image processing, and beyond.

\section{ACKNOWLEDGMENTS}
We acknowledge support from National Science and technology council, Taiwan  under
Grants NSTC 113-2119-M-033-001, by the research project Applications of quantum computing in optimization and finance. We also thank the NVAITC and the NVIDIA Quantum team for their continued technical support. Our thanks extend to the NVIDIA Academic Grant Program for providing access to the GPU resources necessary for this research, and to the NVIDIA Strategic Researcher Engagement team for their support.


\bibliography{sn-bibliography}% common bib file

\end{document}